# Contact-interference Hybrid lithography: Toward Scalable Fabrication of cross-scale periodic micro structure and demonstration on infrared micro polarizer array

Tianshi Lu[a&], Fuyuan Deng[a&], Yufeng Wei[a], Zhipeng Zeng[a] Xinghui Li⋆[a]

[a]Tsinghua Shenzhen International Graduate School, Tsinghua University,

Shenzhen, 518055, China

⋆Corresponding author: li.xinghui@sz.tsinghua.edu.cn

Abstract：

Subwavelength grating micro-polarizer arrays, as a type of focal plane division simultaneous detection method, are significantly advancing the development and practical application of polarization imaging technology. Based on the cross-scale, dual-period characteristics of the grating array, this paper proposes a fabrication method that combines laser interference lithography with contact lithography. This method constructs a complete single-layer micro-polarizer array photoresist pattern through a four-step lithography process. Compared to traditional point-by-point fabrication methods like EBL and FIB, the patterning time is reduced by 3 to 4 orders of magnitude. Additionally, by introducing a refractive index matching liquid and an alignment method based on substrate contours, the effects of gaps and splicing errors are minimized, resulting in high-quality photoresist patterns with splicing errors less than 1 μm. Finally, a double-layer metal grating structure was obtained through pattern transfer. Performance tests show that the micro-polarizer array achieves a maximum transmittance of over 50% and an extinction ratio exceeding 15dB in the 3-15 μm wavelength range. This exploration offers a low-cost, high-efficiency path for fabricating micro-polarizer arrays.

Keywords：Micro-polarizer array, laser interference lithography, mask, lithography

alignment, cross-scale, dual-period

Introduction

In classical field theory, light is considered an electromagnetic wave carrying information such as frequency, amplitude, phase, and polarization. Traditional imaging technology primarily captures frequency and amplitude information, whereas polarization imaging uses the polarization characteristics of light to detect and image targets, obtaining additional information such as the degree of polarization and polarization angle. This broadens the dimensions of target capture, such as surface information of materials and stress information in transparent materials. In atmospheric spectra, the mid-long wave (mid-infrared 3~5μm and far-infrared 8~14μm) lies within the atmospheric window, exhibiting high transmittance. Compared to visible light, infrared waves are less affected by Rayleigh scattering from gas molecules, reducing information loss during atmospheric transmission. Combining polarization information, infrared polarization detection can achieve target detection under conditions of small temperature differences and high-brightness backgrounds. Hence, infrared polarization imaging technology is often employed in the remote sensing field to detect targets.

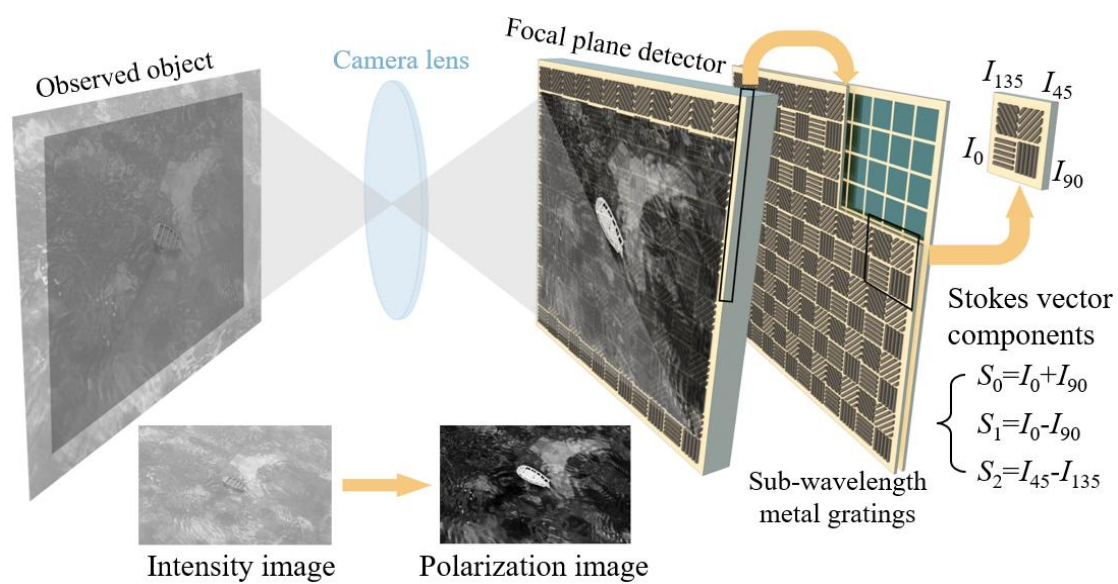

Figure 1 The Application Scenarios of Wavefront-Splitting Polarization Grating Arrays

The periodic units of the micro-polarizer array integrate wire-grid polarizers with different orientations, forming a metal grating with a period smaller than the wavelength of the incident light in the medium. However, the fabrication process of micro-polarizer arrays differs from that of traditional grating manufacturing processes. Arranging subwavelength metal wire grids into anisotropic pixel forms is a significant challenge. The fabrication of micro-polarizer arrays involves micro-nano patterning techniques, including projection lithography, contact lithography, electron beam lithography (EBL), nanoimprint lithography, and laser interference lithography. From the perspective of implementation, electron beam and ion beam direct writing processes create high-resolution patterns point by point [8-10], belonging to point-by-point fabrication techniques. Projection lithography, contact lithography, and nanoimprint lithography rely on the pattern transfer of high-precision masks or molds [7], where the masks or molds often require point-by-point fabrication processes as well, categorizing them as indirect pattern transfer techniques. On the other hand, laser interference lithography uses the interference field formed by light beams to create large-area, tunable grating patterns [11-13], falling under the category of spatial light field transfer techniques. The first two methods require setting the processing resolution according to the smallest feature size of the pattern, allowing the fabrication of arbitrary micro-nano structures within this minimum scale, offering high degrees of freedom at the cost of efficiency and economic viability. In contrast, laser interference lithography, despite its lower patterning freedom, can easily produce periodic patterns with the advantages of low system complexity and high efficiency.

In the existing lithographic pattern fabrication methods for micro-polarizer arrays, electron beam direct writing is challenging for wafer-scale manufacturing, while projection lithography faces strict depth-of-focus limitations. Existing methods based on laser interference lithography often require multi-layer fabrication, increasing the metal layer thickness and making it difficult to ensure consistent polarization performance. These methods also involve multiple lithography, etching, and other processes, leading to considerable complexity [14-16].

To reduce process complexity and improve the fabrication efficiency, this study proposes a cross-scale fabrication method to construct complete patterns on a single layer of photoresist. The micro-polarizer array exhibits a distinctive dual-period characteristic, involving a two-dimensional array period and a periodic grating structure within a single array unit. The scales of the two periods differ significantly, which can be classified into the physical effect scale of subwavelength gratings for achieving polarization selection effects and the engineering structure scale of the polarizer array to match the detector array, as shown in Figure 2. Therefore, since this structure is not a simple single-period structure, traditional mechanical scribing, laser interference lithography, and other methods cannot be directly applied to the processing of this structure. EBL and other point-by-point fabrication methods offer far more degrees of freedom than required, leading to extremely low processing efficiency. Consequently, this study explores a cross-scale processing method based on laser interference lithography and a mask with engineering structure scale resolution. The interference fringes of light are used to process the physical effect scale structures, and the interference wavefront is segmented using a mask to step-by-step process the grating structures in different quadrants of the array, achieving efficient fabrication of this dual-period pattern structure. By transferring the pattern, micro-polarizers with excellent performance for infrared bands were fabricated. The fabricated photoresist pattern quadrant units had a splicing error of less than 1 μm between units, and a maximum extinction ratio of over 20dB was achieved in the 3-15 μm operating wavelength range.

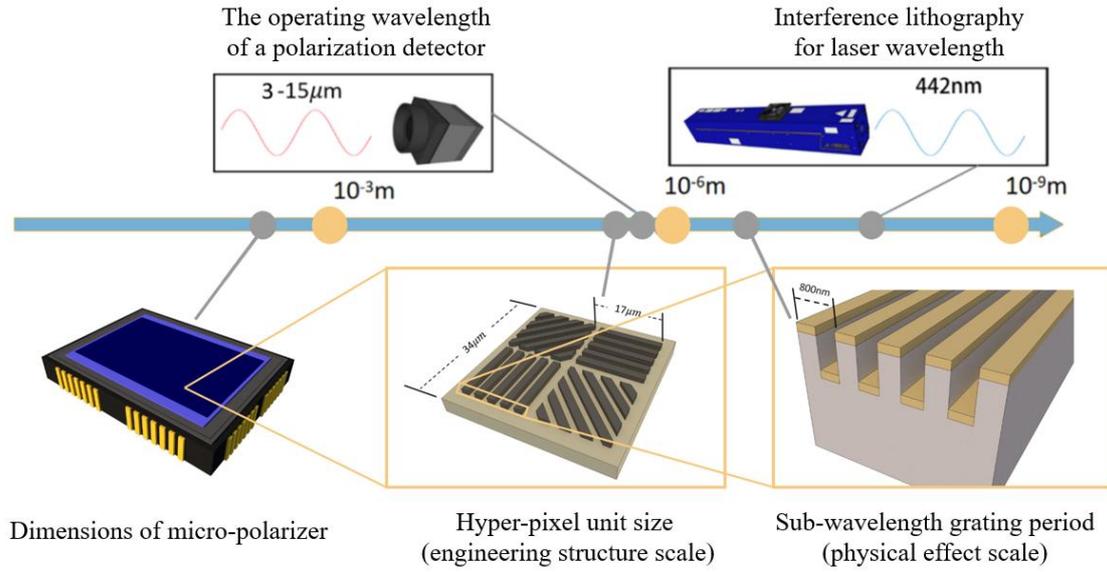

Figure 2 Diagram of Scales Related to the Polarizer

一、 principle

## 2.1 The Stokes Vector Representation of Polarized Light

The amplitude and polarization of light can be represented by the Stokes vector, which uses four parameters, $S_0$、$S_1$、$S_2$、$S_3$, to describe the state of polarization:

$$\begin{cases} S_0 = \langle |E_x|^2 \rangle + \langle |E_y|^2 \rangle = I_0 + I_{90} \\ S_1 = \langle |E_x|^2 \rangle - \langle |E_y|^2 \rangle = I_0 - I_{90} \\ S_2 = \langle 2E_x E_y \cos\delta \rangle = I_{45} - I_{135} \\ S_3 = \langle 2E_x E_y \sin\delta \rangle = I_L - I_R \end{cases} \quad (1)$$

$S_0$ represents the total light intensity, which is the sum of the horizontally polarized component $I_0$ and the vertically polarized component $I_{90}$. $S_1$ represents the difference in intensity between the horizontal polarized component $I_0$ and the vertical polarized component $I_{90}$. $S_2$ represents the difference in intensity between the 45° polarized component $I_{45}$ and the 135° polarized component $I_{135}$. S3 represents the difference between the left-circularly polarized component $I_L$ and the right-circularly polarized component $I_R$. To obtain the Stokes parameters related to polarization, light intensities at angles of 0°, 45°, 90°, and 135° must be acquired. Since circular polarization information is scarce in nature and subwavelength metal gratings, such as wire grid structures, often lack the capability to detect circular polarization information in a single region, the first three Stokes components, $S_0$, $S_1$, and $S_2$, are typically used to calculate the degree of linear polarization and the polarization angle, which are then used to generate polarization images. Therefore, to obtain the light intensities at 0°, 45°, 90°, and 135°, the structure of micro-polarizer arrays is usually designed such that each quadrant unit on the chip contains metal gratings oriented at 0°, 45°, 90°, and 135°.

**2.2 Effective Medium Theory of Subwavelength Gratings**

When the period of a one-dimensional grating is much smaller than the wavelength of the incident light, it can be regarded as a uniform anisotropic dielectric material. Its dielectric constant can be represented by a complex symmetric second-order tensor:

$$\varepsilon_{ij} = \begin{bmatrix} \varepsilon_{11} & \varepsilon_{12} & \varepsilon_{13} \\ \varepsilon_{21} & \varepsilon_{22} & \varepsilon_{23} \\ \varepsilon_{31} & \varepsilon_{32} & \varepsilon_{33} \end{bmatrix} \quad (2)$$

Since the subwavelength grating structure satisfies mirror symmetry in three demonsions, its dielectric tensor satisfies: $\varepsilon_{ij} = A_x \varepsilon_{ij} A_x^T = A_y \varepsilon_{ij} A_y^T = A_z \varepsilon_{ij} A_z^T$

$$(3)$$

Where $A_x = \begin{bmatrix} -1 & 0 & 0 \\ 0 & 1 & 0 \\ 0 & 0 & 1 \end{bmatrix}$, $A_y = \begin{bmatrix} 1 & 0 & 0 \\ 0 & -1 & 0 \\ 0 & 0 & 1 \end{bmatrix}$, $A_z = \begin{bmatrix} 1 & 0 & 0 \\ 0 & 1 & 0 \\ 0 & 0 & -1 \end{bmatrix}$

Substituting into the above equation yields the dielectric tensor for the subwavelength grating as a diagonal matrix:

$$\varepsilon_{ij} = \begin{bmatrix} \varepsilon_{11} & 0 & 0 \\ 0 & \varepsilon_{22} & 0 \\ 0 & 0 & \varepsilon_{33} \end{bmatrix} \quad (4)$$

suppose the incident light is normal, its polarization vector is $[E_x, E_y, 0]^T$, and its transmission properties are only related to the dielectric components $\varepsilon_{11}$, $\varepsilon_{22}$.

The spatial superposition relationship of dielectric constants is given by:

$$\varepsilon_{11} = (1-\eta)\varepsilon_1 + \eta\varepsilon_g \quad (5)$$

$$\varepsilon_{22} = [(1-\eta)\varepsilon_1^{-1} + \eta\varepsilon_g^{-1}]^{-1}$$

For non-magnetic grating materials, their complex refractive index satisfies:

$n_{ij} = \sqrt{\varepsilon_{ij}}$. Therefore, the polarization characteristics of metal wire gratings can be analyzed using effective medium theory. For TM and TE polarized light (perpendicular or parallel to the grating lines), the effective refractive index of the device is:

$$\begin{cases} n_{TM} = \dfrac{(n_1+ik_1)(n_2+ik_2)}{\sqrt{(1-f)(n_2+ik_2)^2 + f(n_1+ik_1)^2}} \\ n_{TE} = \sqrt{(1-f)(n_1+ik_1)^2 + f(n_2+ik_2)^2} \end{cases} \quad (6)$$

Where $n_1$ and $n_2$ represent the refractive indices of the medium between the grating lines and the grating lines themselves, $k_1$ and $k_2$ are their corresponding extinction coefficients, and $f$ is the duty cycle. If the medium between the metal gratings is considered as a layer of air, its complex refractive index $k_1=0$, and the metal grating is considered a perfect metal conductor with a complex refractive index $k_2 \to \infty$. Substituting into the equation yields:

$$\begin{cases} n_{TM} = \dfrac{n_1}{\sqrt{(1-f)}} \\ n_{TE} = ik_2\sqrt{f} \end{cases} \quad (7)$$

Numerically, for TM polarized light, the effective refractive index is a real number, and the subwavelength grating is equivalent to a transparent dielectric layer; for TE polarized light, the effective refractive index is an imaginary number, and the grating is equivalent to a reflective or absorptive metal layer. Thus, subwavelength metal gratings exhibit polarization selectivity when exposed to incident light from different polarization directions. For polarization devices, transmittance and extinction ratio are commonly used to measure their polarization performance. Transmittance represents the device's utilization of light energy, while the extinction ratio characterizes its selective transmission ability.

**2.2 Fabrication Scheme**

In the representative processing techniques for fabricating grating patterns, electron beam lithography (EBL) offers the advantage of high flexibility, allowing exposure based on arbitrarily set patterns. Laser interference lithography, on the other hand, achieves exposure through the periodic interference fringes of dual beams. When comparing the production cost of photoresist patterns, processing a 500×500μm area using EBL typically takes up to 10 hours, whereas a single exposure in laser interference lithography requires only about 50 seconds, with the processing area easily reaching over 20×20mm. Clearly, it has the significant advantages of being able to process large areas with high efficiency. However, its disadvantage is also evident: the processing freedom is lower than required for arrayed grating, as dual-beam interference exposure can only form one-dimensional grating structures.

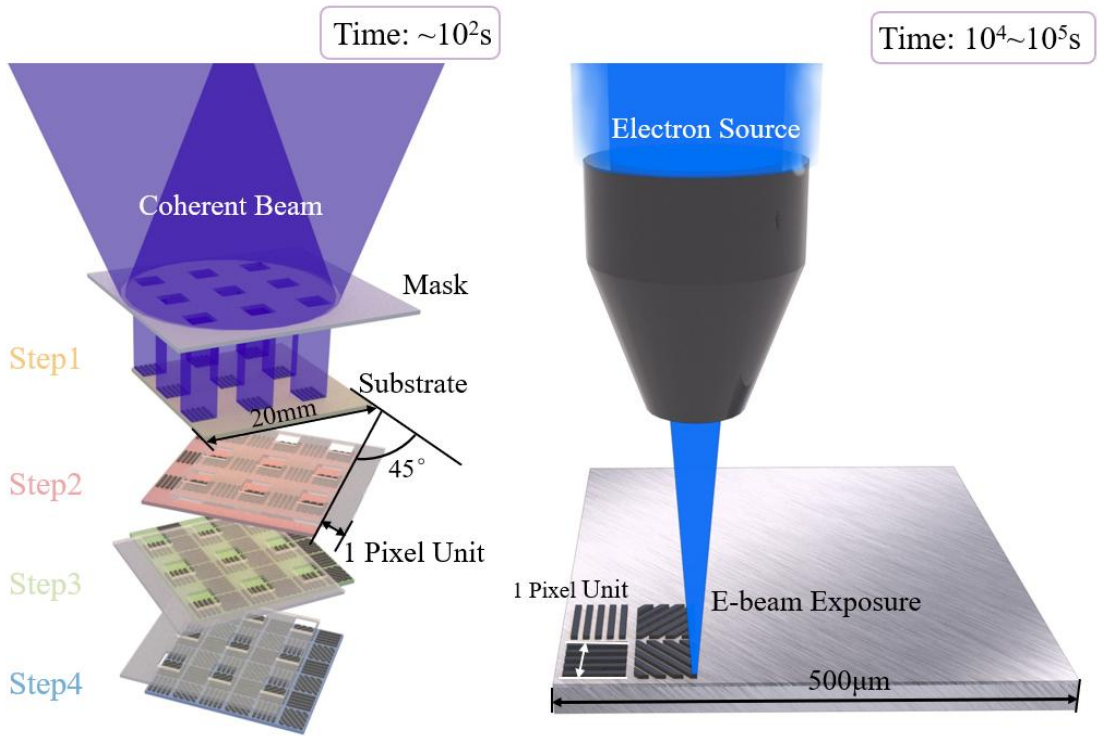

Figure 3 The High-Efficiency Array Processing Method Based on Interference Lithography Proposed in This Study and Its Comparison with Traditional Electron Beam Lithography

Dual-beam laser interference lithography can only form uniformly oriented photoresist gratings across the entire surface. Therefore, this study introduces a mask and employs stepwise contact exposure to stitch patterns together, thereby increasing processing flexibility. On the one hand, the combination of these two lithography methods reduces the pattern resolution of the mask used in contact exposure from the grating line width down to the detector pixel scale. On the other hand, the selective transmission or blocking of certain regions of the mask serves to trim the grating patterns formed by laser interference lithography.

The key to constructing a single-layer photoresist pattern on the substrate lies in the positional adjustments between the mask and the substrate during different steps. By adjusting the relative displacement between the mask and the substrate, the quadrant regions for exposure in different steps are determined; through the overall rotation of the mask and substrate, the grating orientations formed in different steps are defined. By splicing together the patterns formed by four-step interference exposure, a complete micro-polarizer array photoresist pattern is achieved (Figure 3(a)).

二、 **Experiments and Discussion**

### 3.1 Design of material and structure

We used p-type double-sided polished silicon wafers as the substrate material, which have good transmittance in the working wavelength range of 3-15 μm. Aluminum was chosen as the material for the metal grating due to its high extinction coefficient in the infrared range (30-115), which, according to the effective medium theory, provides excellent polarization performance. The micro-polarizer array consists of subwavelength metal gratings oriented at 0°, 45°, 90°, and 135°, capable of acquiring the first three Stokes vector components, thereby enabling the generation of polarization images. The subwavelength metal gratings are located in the four quadrant unit areas of the periodic unit, as shown in Figure 4. The center-to-center distance between adjacent pixels of mainstream infrared detectors is 17 μm, with square pixel dimensions of 15 μm and a 2 μm gap to prevent crosstalk. The center-to-center distance and size of the quadrant units in the micro-polarizer perfectly match those of the detector. Compared to the period, it is challenge to controlling the duty cycle of gratings processed by holographic lithography，since it could affected by laser fluxdensity, exposure time, development time, temperature, etching time and selection ratio and other factors. Inspired by equation (7) and verified by the FDTD simulation algorithm, using a double-layer metal grating with a complementary metal dielectric duty cycle as shown in Figure 4(a) can reduce the impact of the extinction ratio on changes in the duty cycle. The simulation results are shown in Figure 4(b). In addtion,This structural design can avoid etching or lifting off the metal during the photoresist pattern transfer process, thus improving the stability of pattern transfer. For detail, period is set to be 800nm, satisfying the subwavelength condition. The dielectric grating is obtained by etching the substrate silicon to a height of 900nm, and the two metal layers are both 80nm in height.

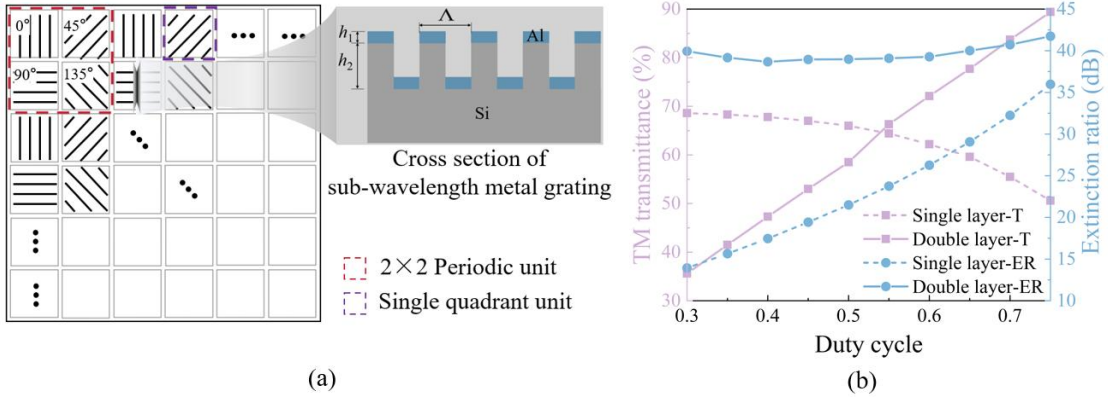

(a)    (b)

Figure 4 Structural Design and Simulation Results of the Polarization Grating Array for the Infrared Band

## 2.3 System Setup and Optimization

To experimentally demonstrate the feasibility of this fabrication scheme, a lithographic pattern processing system was set up, primarily consisting of a single-axis Lloyd's mirror interference exposure optical path and a processing workstation connected to it, as shown in Figure 5. The system uses a helium-cadmium laser as the light source, with a wavelength $\lambda$ of 441.6 nm. The resulting grating period g is calculated using Equation 8.

$$g = \frac{\lambda}{2\sin\theta} \quad (8)$$

Where $\theta$ represents the angle between the interference light incident on the substrate and the normal, which can be accurately controlled by adjusting mirrors 1-2. The light emitted from the laser first passes through a beam-expanding collimating unit to form a wide, uniformly energized beam. This beam-expanding collimating unit mainly includes a 40×, NA=0.65 objective lens, a 10μm diameter pinhole, an adjustable aperture, and a collimating lens with an effective focal length of 500mm.

The beam then enters the Lloyd's mirror above the workstation at an oblique angle by adjusting the deflection of mirrors 1-2, achieving wavefront-splitting interference exposure. The workstation uses a clamping structure and suction cups to securely fix the mask and substrate, ensuring that their positions remain unchanged during exposure. It also has four degrees of freedom of adjustment (x, y, z, and rotation around the z-axis), with a minimum repeatable incremental movement of 0.2 μm, enabling relative positional adjustments between the mask and substrate in the system. The system's exposure time is controlled by a high-response-speed shutter.

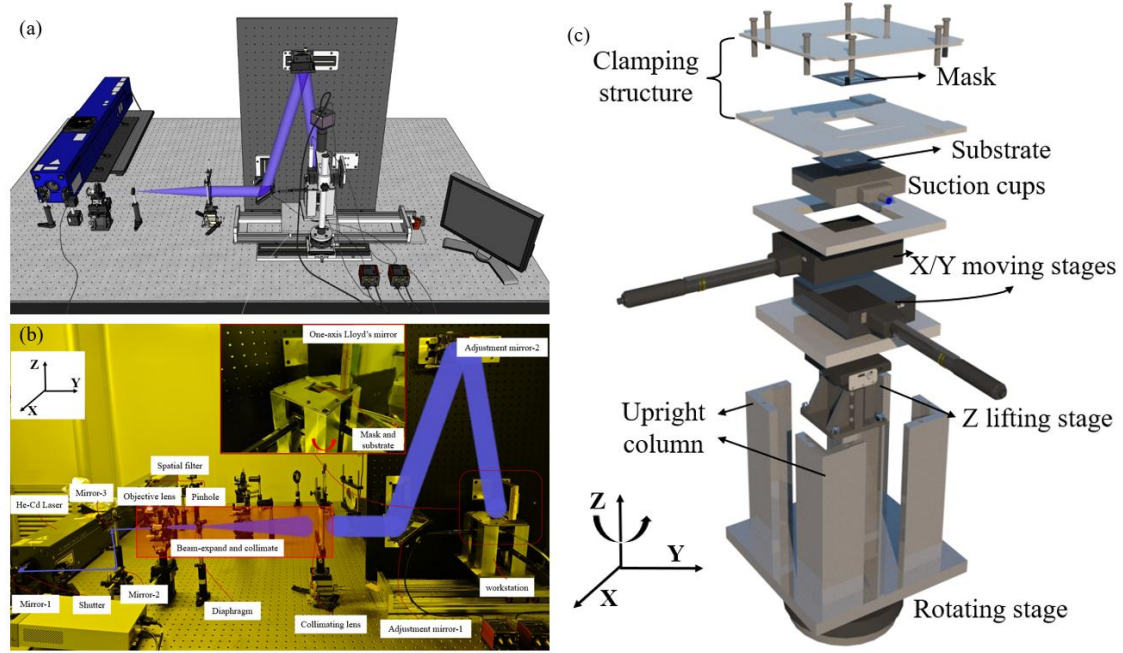

Figure 5 (a) the model of the laser interference lithography system, (b) the real view of the system, (c) the structure of the mask-substrate alignment system.

As shown in Figure 6, the mask is placed in contact with the exposed substrate in the system. The mask mainly includes a central processing area and ruler scale areas on both sides. The central processing area is 20mm in length and contains approximately 600×600 periodic units. The light-transmitting window is arranged in the upper-left corner of the periodic unit to receive interference exposure and form grating lines in a single step. The mark area is composed of solid and dashed lines, serving as distinguishable alignment references between different steps to achieve pattern splicing between quadrants. The light-transmitting window spacing and solid-dashed line spacing of the mask are determined by the quadrant unit spacing of the micro-polarizer array to be fabricated (i.e., the pixel size), designed as 34 μm and 17 μm.

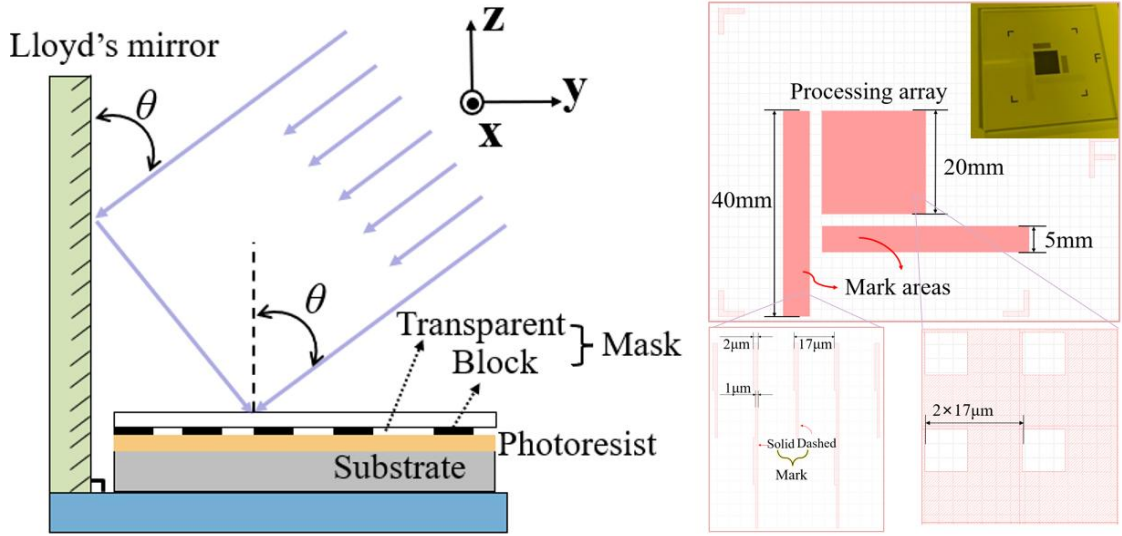

Figure 6 The design of a mask with alignment functionality for merging proximity lithography and interference lithography.

In the experiments, we found that the fabricated photoresist pattern exhibited large periodic macro-stripe effects, causing inconsistent grating groove depths in different array units of the micro-polarizer, thereby reducing the pattern quality. Factors such as the misalignment between the substrate and the mask, and the unevenness of the substrate surface, can lead to the formation of gaps. Here, we analyze the negative impact of these gaps on exposure. First, the incident light undergoes multiple reflections within the gap, and the reflected light interferes with the incident light. Due to the varying thicknesses of the gaps in different areas, this interference results in a non-uniform distribution of brightness, forming macro-stripes, as illustrated in Figure 7(a).

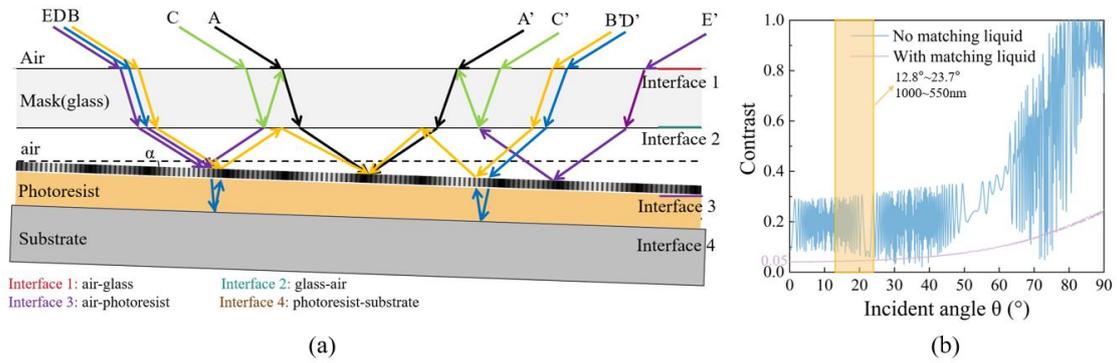

Figure 7 (a) Mechanism of macro-stripe formation; (b) Comparison of stripe contrast with and without matching liquid at all incident angles.

To reduce the contrast of these macro-stripes, we used a refractive index matching liquid to fill the gap, minimizing light reflection between multiple layers of media. Glycerol was selected as the refractive index matching liquid, with a refractive index (1.48) close to that of the mask (glass, 1.53) and photoresist (1.69). Its water-soluble property ensures it does not interfere with subsequent development processes. The simulated contrast of macro-stripes with and without matching liquid is shown in Figure 8(b). As illustrated in Figure 5, the photoresist pattern of the single-direction micro-polarizer is compared before and after adding the matching liquid. In Figure 8(a), it is evident that the stripe conditions vary across different areas, with some areas in the lower-right corner showing missing or faint stripes. In contrast, Figure 8(b) displays relatively uniform and clear stripes with good contrast. Figure 8(c) shows surface observations, where visible macro-stripes disappears when using matching fluid. This indicates that adding a refractive index matching liquid significantly improves the mask quality.

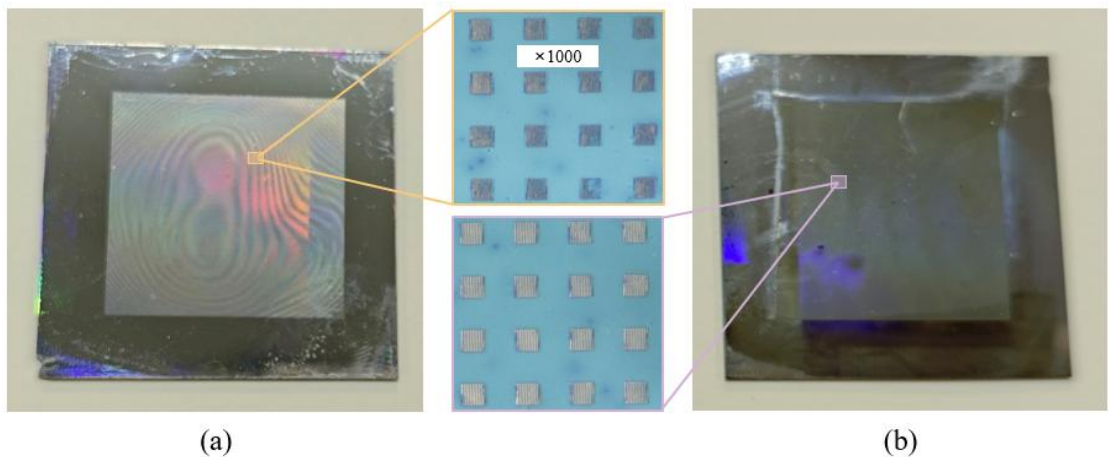

Figure 8 Comparison of single-direction micro-polarizer photoresist pattern results (a) Before adding matching liquid; (b) After adding matching liquid

Additionally, since two beams of coherent light are incident with an angle, the effective interference area (BC segment in Figure 9) decreases as the incident angle and the gap between the mask and substrate increase. Adding matching liquid or reducing the gap can mitigate the loss of processing area due to beam misalignment without changing the period of the interference fringes.

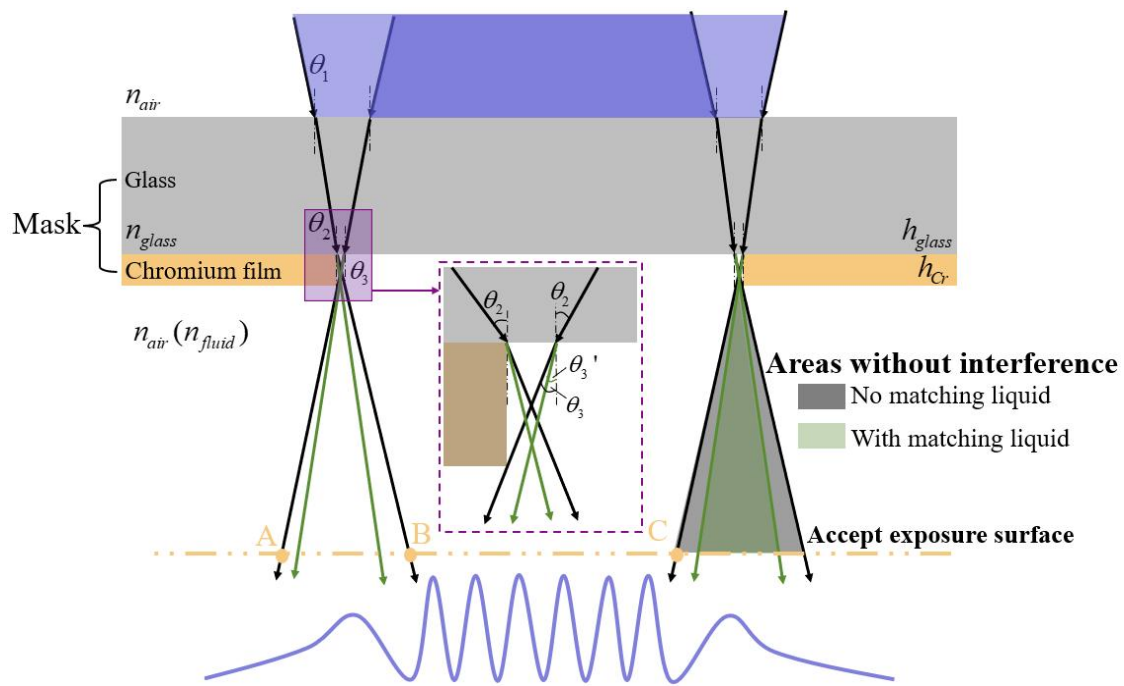

Figure 9 Refraction of light incident on the mask with and without matching liquid.

As analyzed above, the gap between the mask and the silicon substrate significantly affects the imaging quality of the lithography. Therefore, within the system's operational limits, the pressure between the mask and the substrate should be increased as much as possible to reduce the gap between them. Since photoresist is an organic soft material and is more sensitive to pressure compared to glass masks or silicon wafers, it is necessary to explore whether the thickness of the photoresist film will change significantly under a certain pressure range.

We used 20×20mm silicon wafers and applied a uniform coating of photoresist on the surface, resulting in a dried thickness of 140±5nm. We placed weights of 10g, 100g, 1kg, 2kg, and 5kg on top of the mask and maintained the pressure for 30 minutes. We then selected five measurement points at the center and four corners of the silicon wafer to test the thickness of the photoresist film. The results are shown in Figure 10. The results indicate that within a pressure range up to 12500Pa, no significant changes in the thickness of the photoresist film beyond the measurement error range were observed. Since the pressure range tested exceeds the upper limit of pressure due to the load on the translation stage and the mechanical strength of the mask in the actual system, it can be concluded that in this system, active pressure loads only affect the gap between the photoresist and the mask, and have a negligible impact on the thickness of the photoresist film itself.

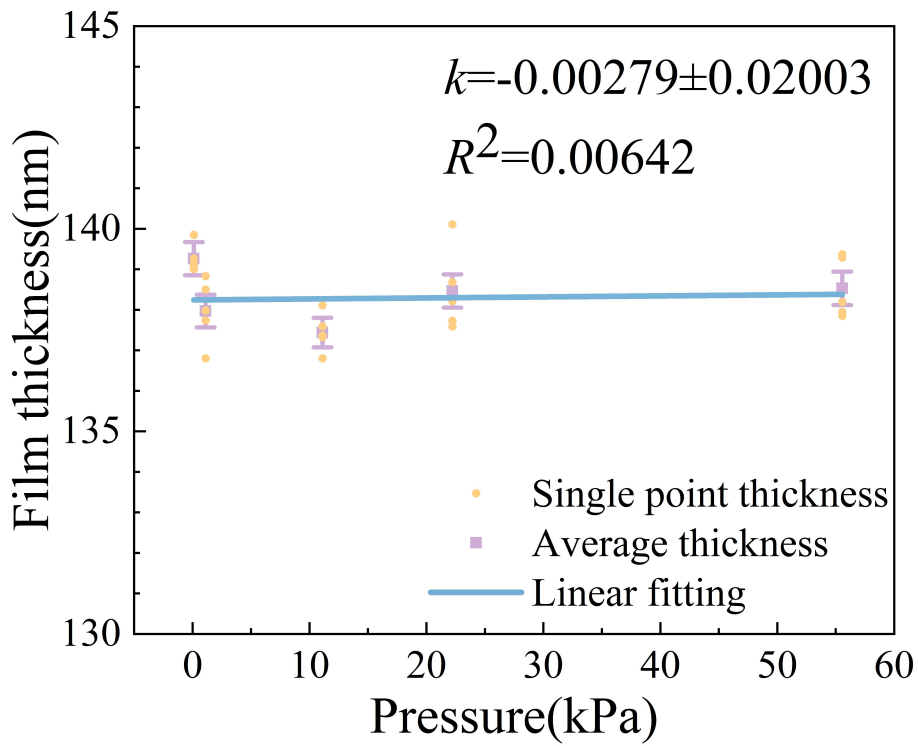

Figure 10 Pressure resistance test of the photoresist.

Additionally, to ensure the splicing quality of patterns in each quadrant, this study proposed a substrate contour-based alignment scheme to assist in achieving high-precision alignment between the mask and substrate. The main alignment strategy is as follows (Figure 11):

Before each exposure, align the contour lines of the substrate with the marks under a microscope, ensuring the entire system is aligned. Record the type of mark (solid or dashed) aligned with the left and bottom contour lines of the silicon wafer at this time. Then, when moving the substrate along the X or Y direction to the exposure area corresponding to the current step, observe the current position of the substrate under the microscope. If moving along the X direction, align the left contour line of the substrate with the opposite type of mark from the previous alignment, while keeping the bottom contour line aligned with the same type of mark. If moving along the Y direction, align the bottom contour line of the substrate with the opposite type of mark from the previous alignment, while keeping the left contour line aligned with the same type of mark.

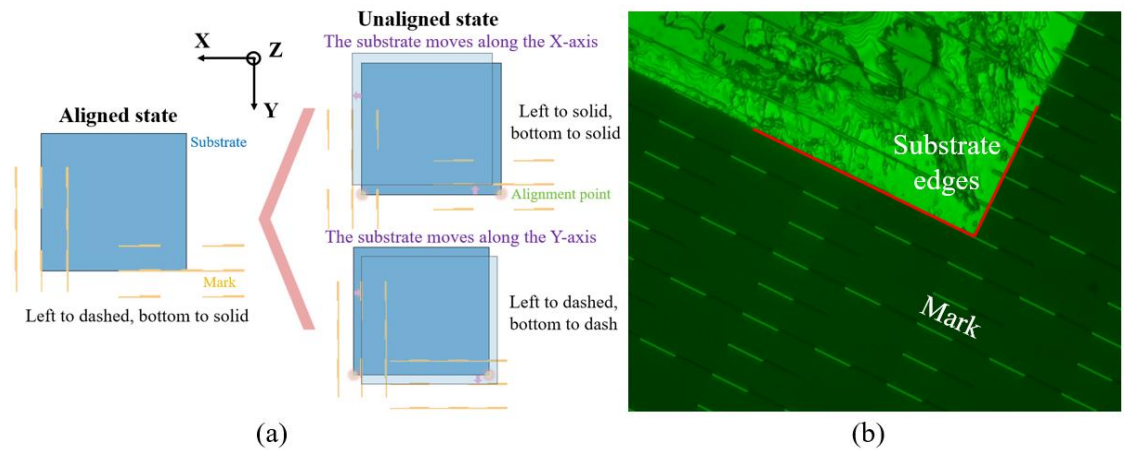

Figure 11 (a) Mask-substrate alignment strategy (b) Actual alignment result.

To achieve this, an optical microscope was installed on one side of the workstation for alignment purposes. The microscope has a 20X magnification, a working distance of 8mm, and is equipped with a 2-megapixel camera. The microscope uses green coaxial epi-illumination, avoiding the wavelength range to which the photoresist is sensitive, allowing the observation of the microstructure on the underside of the glass mask, meeting the requirements of this study for aligning the substrate contour with the marks.

The designed micro-polarizer array has a pixel spacing of 2 μm, and when the splicing error is ≤1 μm, the micro-grating structures between units remain intact without overlap, preventing crosstalk between signals. Even when integrated with a CCD detector, any slight misalignment between the grating units and the detector pixels will only marginally affect transmittance without significantly impacting the extinction ratio.

In the system, the observation accuracy is mainly limited by the optical resolution of the microscope. Since the microscope system used in this study has a pixel size of 3.45 μm×3.45 μm, with a 20× optical magnification, the pixel corresponding size is 0.17 μm. Additionally, with an NA value of 0.45 and using green illumination at a wavelength of 525nm, the system's resolution limit, calculated according to the diffraction limit, is 0.71 μm. The combined observation accuracy after error synthesis is 0.73 μm, and a comprehensive analysis indicates that the alignment error does not exceed 1 μm, meeting the design requirements.

## 2.4 Fabrication

Figure 12(a) shows the process flowchart for fabricating the micro-polarizer.

(1) First, apply photoresist to a clean silicon substrate. The photoresist used is S1805, mixed with PGMEA solvent in a 1:1 ratio. It is spin-coated using a spin coater at 1000 rpm for 3 seconds for pre-coating, followed by 6000 rpm for 30 seconds, and then baked on a hot plate at 105°C for 5 minutes. The result is a uniform thin film of approximately 150nm thickness.

(2) The system shown in Figure 5 was set up to create the photoresist pattern. The Lloyd's mirror was set at an incident angle of 16°, and exposure was carried out at a power density of 1.62mW/cm² for 45 seconds. The pattern was then developed by soaking in ZX-238 developer for 20 seconds, resulting in the photoresist pattern shown in Figure 12. The entire array measures approximately 20×20mm, with over a million pixels. The pattern features clear and straight grating lines with good orientation. The splicing between quadrant units is well-aligned, with a splicing error of less than 1 μm, and the periodic units are consistent.

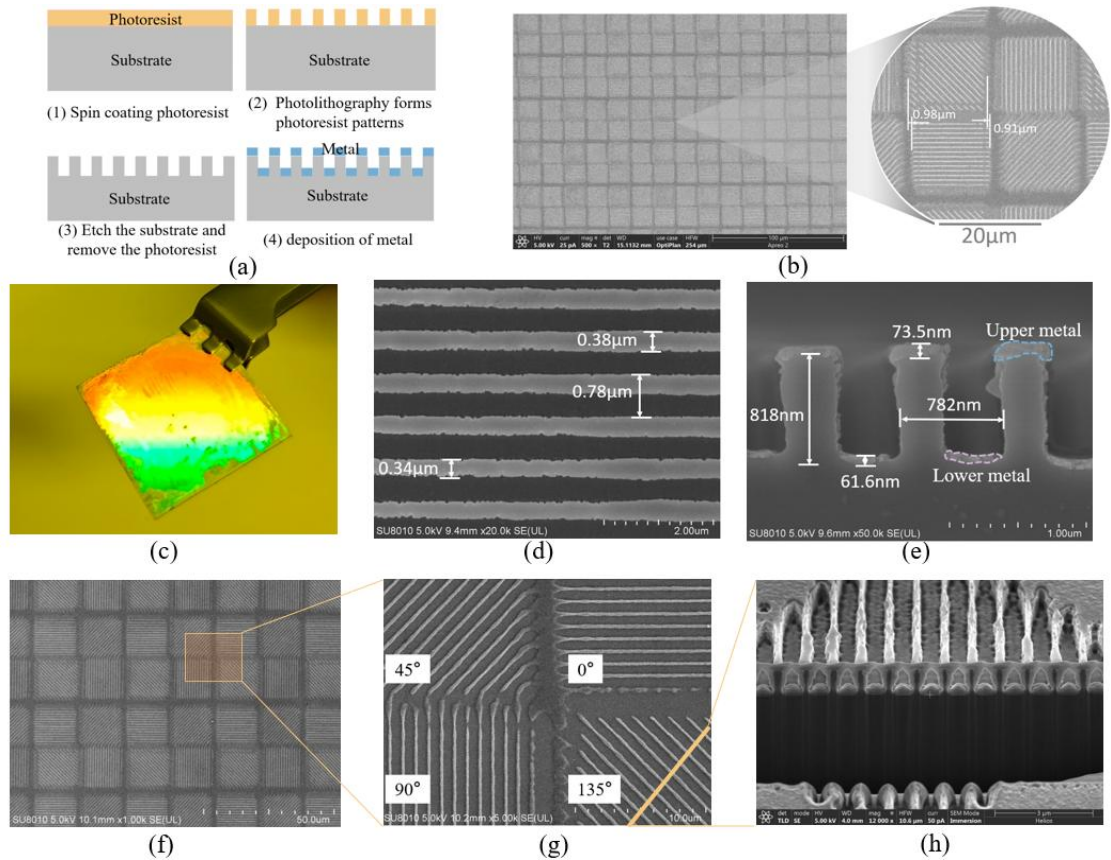

Figure 12 Basic Processing Workflow and Key Step Results for Grating Fabrication: (a) Process Flowchart, (b) SEM Image of Photoresist Pattern, (c) Appearance of the Polarizer, (d)(e) SEM Images of the Double-Layer Subwavelength Metal Grating, (f)-(g) SEM Images of the Micro-Polarizer Structure.

(3) The photoresist pattern was transferred onto the silicon substrate using ICP-RIE (Inductively Coupled Plasma Reactive Ion Etching) to obtain a silicon dielectric grating structure. SF6 was used as the reactive gas with silicon, while O2 was added to increase the etching rate. Argon (Ar), an inert gas that does not participate in the reaction, was used to purify the chamber. The process recipe used in the experiment is shown in Table 1.

Table 1 ICP Etching Process Recipe

| Parameters | Stabilize Chamber | Remove Residual Photoresist | Introduce Reactive Gases | High-Power Etching | Purify Chamber |
|---|---|---|---|---|---|
| Time(s) | 20 | 10 | 10 | 25 | 60 |
| Pressure(mTorr) | 8 | 8 | 30 | 30 | 50 |
| Ar(500sccm) | 0 | 0 | 0 | 0 | 200 |
| $O_2$(300sccm) | 50 | 50 | 40 | 40 | 0 |
| $SF_6$(200sccm) | 0 | 0 | 90 | 90 | 0 |
| Source Power(W) | 0 | 100 | 0 | 600 | 0 |
| Bias Power(W) | 0 | 20 | 0 | 30 | 0 |

（4）Using electron beam evaporation, 80nm of aluminum was vertically deposited onto the sample at a rate of 0.7 Å/s, forming uniform and dense metal aluminum layers on both the upper and lower surfaces of the silicon dielectric grating. To facilitate subsequent performance testing, a full-surface single-direction polarization grating was also fabricated. Figure 12 shows the SEM images of the fabricated results, which clearly retain the periodic structure with good morphological characteristics.

### 2.5 Optical performance Testing

In this study, the performance of the fabricated polarization devices was tested using a Fourier transform infrared spectrometer (Nicolet IS50R). This instrument can generate infrared bands covering the wavelength range of 3-15 μm, allowing for adjustable polarization states of the light source and collecting transmittance information after passing through the sample. The extinction ratio was calculated using Equation 9.

$$ER = 10\log\frac{T_{TM}}{T_{TE}} \quad (9)$$

For the performance testing of the micro-polarizer array, a masking structure was added to measure the performance of the single-direction array. To minimize the impact of this structure on the test, it needed to be as thin as possible with minimal gaps between it and the sample, though such structures are often difficult to manufacture independently. A reasonable equivalent testing method is shown in Figure 13(d).

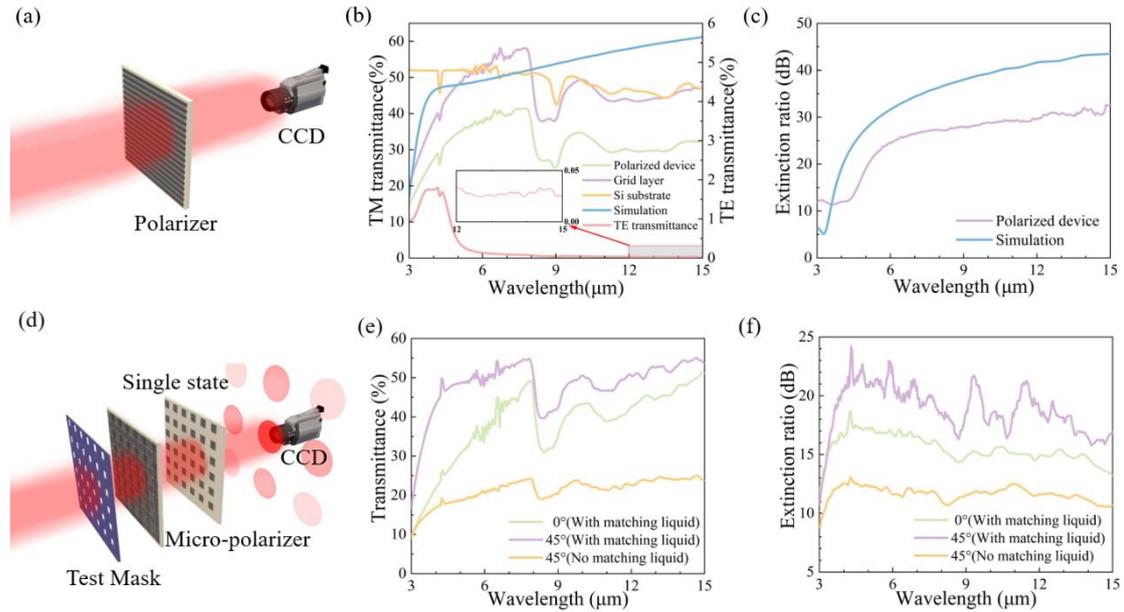

Figure 13 the micro-polarizer performance testing setup and results: (a) Testing of a full-surface single-direction polarization grating; (b) Transmittance; (c) Extinction ratio; (d) Equivalent testing for the micro-polarizer array; (e) Comparison of transmittance of the micro-polarizer array with and without refractive index matching liquid; (f) Comparison of extinction ratio of the micro-polarizer array with and without refractive index matching liquid.

The testing results for the full-surface single-direction polarization grating are shown in Figures 13(b) and (c). After removing substrate reflection and attenuation, the transmittance curve of the grating layer closely matches the simulation results, with the maximum TM transmittance exceeding 50% (58.2%) in certain bands. The enlarged view shows that the TE transmittance of the polarizer decreases to approximately 0.025% in the wavelength range above 12 μm, demonstrating strong shielding for TE light. The extinction ratio exceeds 30dB (33.8dB) and remains above 20dB for wavelengths above 5 μm, proving the superior polarization performance of the double-layer metal structure.

Among the different orientations of the micro-polarizer array, the unit arrays in the 0° and 90°, and 45° and 135° directions are equivalent in testing. Therefore, this study only tested the performance of the 0° and 45° directions. The testing results are shown in Figures 13(e) and (f). The results indicate that the micro-polarizer array's maximum extinction ratio exceeds 20dB (55.0%, 24.2dB). Additionally, the performance test results of samples without the matching liquid showed that the extinction ratio decreased by approximately 6dB.

These testing results differ from those of the full-surface single-direction polarization grating. In terms of transmittance, the transmittance obtained using this testing method is a weighted result of two regions illuminated by the light spot—namely, the aluminum film portion and the grating portion. Therefore, the total transmitted energy is reduced. Additionally, the single-direction micro-polarizer array can be considered equivalent to a two-dimensional grating with a period of 34μm. When the light spot passes through this structure, diffraction occurs, and the light reaching the detector may only include the zero-order diffraction light, further reducing the total energy (as shown in Figure 13(d)). However, since this method involves far-field testing, the impact introduced by this structure should be excluded when calculating grating transmittance. As for the extinction ratio, the value obtained under this structure is a weighted average of the two regions, with the presence of the aluminum film leading to a decrease in the extinction ratio. Additionally, the limited number of stripes in the single-direction micro-polarizer array, along with the aspect ratio of the grating structure being far less than that of the full-surface single-direction grating, also contributes to the decrease in extinction ratio.

**Conclusion**

This study focuses on the design and fabrication of micro-polarizer arrays. Subwavelength gratings were processed using laser interference lithography to achieve the device's fundamental physical functions, combined with contact lithography to arrange the subwavelength gratings in an array, enabling the zonal fabrication of photoresist grating patterns with different orientations. Compared to point-by-point fabrication processes such as electron beam direct writing, this method requires only four short exposures, one non-metallic etching step, and one thin-film metal deposition, significantly reducing the complexity of the fabrication process. This method offers high efficiency, low cost, and has the potential for mass production. The processing system designed and constructed according to the fabrication scheme uses a refractive index matching liquid to compensate for gaps and a substrate contour-based alignment method to stably produce high-quality micro-polarizer photoresist patterns. SEM characterization results show that the splicing error is less than 1 μm. After subsequent pattern transfer processes, a double-layer metal grating structure was formed, resulting in a micro-polarizer array. Testing showed that in the working wavelength range of 3-15 μm, the maximum transmittance exceeded 50%, and the maximum extinction ratio was over 20dB, demonstrating superior polarization performance that meets practical application requirements. The results show that the cross-scale processing method proposed in this study offers significant efficiency advantages for processing structures with dual-period characteristics compared to traditional methods, and the method could be extended to other similar surface optical functional structures, such as metasurfaces and arrayed diffractive devices.